\documentclass[a4paper,11pt]{article}
\pdfoutput=1 

\usepackage{jheppub} 
\usepackage{graphicx}
\usepackage{amssymb,amsmath}

\title{\Large{\bf Attractive Interaction between Vortex and Anti-vortex in Holographic Superfluid } }

\author[a,b]{Shan-Quan Lan,}
\author[a,b]{Gu-Qiang Li,}
\author[a,b]{Jie-Xiong Mo,}
\author[a,b]{Xiao-Bao Xu}

\affiliation[a]{Department of Physics, Lingnan Normal University, Zhanjiang, 524048, Guangdong, China}
\affiliation[b]{Institute of Theoretical Physics, Lingnan Normal University, Zhanjiang, 524048, Guangdong, China}

\emailAdd{shanquanlan@126.com}

\abstract{
Annihilation of vortex and anti-vortex in two dimensional turbulent superfluid are important phenomena which reduce the topological defects. In this paper, we report new findings on the annihilation process of a pair of vortices in holographic superfluid. The process is found to consist of two stages which are amazingly separated by vortex size $2r$. The separation distance $\delta(t)$ between vortex and anti-vortex as a function of time is well fitted by $\alpha (t_{0}-t)^{n}$, where the scaling exponent $n=1/2$ for $\delta (t)>2r$, and $n=2/5$ for $\delta(t)<2r$.  Thus the attractive force between vortex and anti-vortex is derived as $f(\delta)\propto 1/\delta^{3}$ for the first stage, and $f(\delta)\propto 1/\delta^{4}$ for the second stage. Successfully, we present physical interpretation for the theorem that the annihilation rate of vortices in turbulent superfluid obeys the two-body decay law when the vortex density is low.}

\begin{document}
\maketitle
\flushbottom

\section{Introduction}

Turbulence can be roughly defined as a spatially and temporally complex state
of fluid motion\cite{2014PNASbarenhi}. It can be found everywhere, such as rapid stream, smog, airflow, superfluid helium of heat convection, cold atoms being stirred, etc. Meanwhile, as the interaction is nonlinear and involve many length and time scales, turbulence become one of the biggest problems in modern science. Both the classical turbulence and quantum turbulence are  popular and active research directions. Specially, due to the great development of cooling and controlling techniques in recent decades, quantum turbulence has established itself in the turbulence community, shedding new light on both the quantum and classical aspects of turbulence.

Vortices are essential objects in turbulence. They play a crucial role in various phenomena. In quantum turbulence, such as superfluid helium, atomic Bose-Einstein condensates, vortices are topological defects which are quantized. Reconnection of vortex lines in three dimensional superfluid, or annihilation of vortex and anti-vortex in two dimensional superfluid are  important phenomena which reduce the topological defects, dissipate energy and randomise the velocity field\cite{2011jltpbarenhi}. During the reconnection process, the separation distance $\delta (t)$ between two vortex lines varying with respect to time is intensively studied. It is firstly reported by de Waele and Aarts\cite{1994prlwaele} that
\begin{equation}
  \delta(t)=(\kappa/2\pi)^{1/2}\sqrt{t_{0}-t},
\end{equation}
where $\kappa$ is the circulation quantum and $t_{0}$ is the reconnection time. This numerical work is based on Schwarz's vortex filament model\cite{1988prbschwarz}, which assumes a very small vortex core. Thus the above result breaks down when $\delta (t)$ is smaller than the vortex size. Anyway, this $(t_{0}-t)^{1/2}$ scaling is confirmed in He II experiment\cite{2008PNASbewley,2008pdnppaoletti}, and by an approximate analytic solution of the Gross-Pitaevskii Equation (GPE)\cite{2003jltsergey}. However, there are also many numerical studies report similar fitting functions with modified scaling exponents. For example, the fitting functions
\begin{equation}
   \delta(t)=A_{1}(t_{0}-t)^{A_{2}},
\end{equation}
and
\begin{equation}
   \delta(t)=B_{1}(t_{0}-t)^{1/2}[1+B_{2}(t_{0}-t)]
\end{equation}
were reported in Refs.\cite{2011jltpbarenhi,2012prbbaggaley, 2012pfzuccher, 2014praallen,2017prfalberto}, where the scaling exponent $A_{2}$ was found to be varying around $1/2$ and $B_{2}$ is small.

All the above studies are about three-dimensional situations. The two-dimensional cases, such as the oblate Bose-Einstein condensates where vortices are annihilated rather than reconnected,  are also important and interesting. Do they have similar behaviors and scaling exponents? In this paper, we try to solve this problem. We numerically simulate and study the annihilation process of vortex and anti-vortex based on the holographic duality rather than GPE. Holographic duality\cite{Maldacena1998top,gubser1998gauge,witten1998anti} is an alternative theoretical framework to deal with strongly coupled quantum many-body systems which is encoded in a classical gravitational system with one extra dimension. Holographic superfluid model was established in Refs.\cite{20083h1,20083h2}, while soliton and vortex solutions were studied in Refs.\cite{2010prdkeranen1,2010prdkeranen2,2017prdlan}. Recently, holographic superfluid turbulence were studied in Refs.\cite{chesler2013holographic,Ewerz2014tua,du2014holographic,2016jheplan}. Holographic superfluid model is non-perturbational, thus allowing a first-principles investigation of the annihilation process. What's more, it handles  finite temperature cases naturally. While the GPE mainly handles the zero temperature case. Manually reconstructed GPE which involves dissipative effect can handle the finite temperature cases\cite{1958pitaevskii}.

This paper is organised as follows. Holographic superfluid model and relevant numerics are introduced in Sec.\ref{sec2}. Numerical results and attractive interaction between vortex and anti-vortex are analysed and discussed in Sec.\ref{sec3}. Sec.\ref{sec4} devotes to  conclusions and  suggestions on future directions.

\section{set up}
\label{sec2}

For the two dimensional superfluid, a simple holographic model is a gravitational system in asymptotically AdS$_{4}$ spacetime. The corresponding bulk action can be written as\cite{20083h1,20083h2}
\begin{equation}
S=\frac{1}{16\pi{G}}\int_{\mathcal{M}}\mathnormal{d}^{4}x\sqrt{-g}(R+\frac{6}{L^{2}}+\frac{1}{q^{2}}\mathcal{L}_{matter}),
\end{equation}
where the matter Lagrangian reads
\begin{equation}
\mathcal{L}_{matter}=-\frac{1}{4}F_{ab}F^{ab}-|D\Psi|^{2}-m^{2}|\Psi|^{2}.
\end{equation}
Here $D=\nabla-iA$ with $\nabla$ as the covariant derivative compatible to the metric. $A_{a}$  is a dynamical U(1) gauge field and $\Psi$ is a complex scalar field  with mass $m$ and charge $q$.  To simplify the problem, one usually work in the probe limit, which decouples the matter fields from gravity. Thus the Schwarzschild black brane can be written in the infalling Eddington coordinates as
\begin{equation}
ds^{2}=\frac{L^{2}}{z^{2}}(-f(z)dt^{2}-2dt dz+dx^{2}+dy^{2}),
\end{equation}
where the factor $f(z)=1-(\frac{z}{z_{h}})^{3}$ with $z=z_{h}$  corresponding to the horizon and $z=0$ corresponding to the AdS boundary.
Then the equations of motion for the matter fields can be written as
\begin{equation}
D_{a}D^{a}\Psi-m^{2}\Psi=0,\nabla_{a}F^{ab}=i(\overline{\Psi}D^{b}\Psi-\Psi\overline{D^{b}\Psi}).
\end{equation}
In what follows, we will take the units in which $L=1,16\pi Gq^{2}=1$, and $z_{h}=1$, and work with $m^{2}=-2$ in the standard quantization case\cite{breitenlohner1982}. In the axial gauge $A_{z}=0$, the asymptotic solutions of $A$ and $\Psi$ near the AdS boundary can be expanded as
\begin{equation}\label{asymp}
A_{\mu}=a_{\mu}+b_{\mu}z+o(z),\Psi=z[\beta+\psi{z}+o(z)].
\end{equation}
According to the holographic dictionary, the temperature, the expectation values of the conserved current $j^{\mu}$ and the condensate operator $O$ in the superfluid are given by\cite{2013jhepli}
\begin{equation}
T=\frac{3}{4\pi},
\end{equation}
\begin{eqnarray}\label{current}
\langle j^{\mu}\rangle=\frac{\delta{S_{ren}}}{\delta{a_{\mu}}}=\lim_{z\rightarrow 0}\sqrt{-g}F^{z\mu},
\end{eqnarray}
\begin{eqnarray}\label{operator}
\langle O\rangle&=&\frac{\delta{S_{ren}}}{\delta{\beta}}=-\lim_{z\rightarrow 0}z\sqrt{-\gamma}(n_a\overline{D^a\Psi}+\overline{\Psi})\nonumber\\
&=&\overline{\psi}-\dot{\overline{\beta}}-ia_{t}\overline{\beta},
\end{eqnarray}
where the dot denotes the time derivative,and the renormalized action is given by
\begin{equation}
S_{ren}=S-\int_{\mathcal{B}}\sqrt{-\gamma}|\Psi|^{2}
\end{equation}
with the counter term added to make the original action finite.

Here we switch off the sources of the operators by setting
\begin{eqnarray}
a_{x}=0,a_{y}=0,\beta=0.
\end{eqnarray}
Then the superfluid velocity is defined as
\begin{eqnarray}\label{velocity}
\boldsymbol{u}=\frac{\mathcal{J}}{|\psi|^{2}},\mathcal{J}=\frac{i}{2}(\overline{\psi}\boldsymbol{\partial}\psi-\psi\boldsymbol{\partial}\overline{\psi}),
\end{eqnarray}
and the winding number $\sigma$ of a vortex is
\begin{equation}\label{omega}
\sigma=\frac{1}{2\pi}\oint_{c}d\boldsymbol{x}\cdot\boldsymbol{u},
\end{equation}
where $c$ denotes a counterclockwise oriented path surrounding a single vortex. In what follows we determine the position of the vortex by calculating the winding number of each point in the system.

To investigate the annihilation process of a vortex pair with winding number $\sigma=\pm 1$, we consider a periodic $30\times30$ square box on which a pair of vortices evolve freely.  The initial position of the vortices are randomly placed. What's more, a random perturbation velocity field are added to the system. As a result, the vortices have random initial velocities. By defining a new function $\Phi=\frac{\Psi}{z}$, the later time behavior of the system is determined by the following rewritten equations of motion
\begin{eqnarray}\label{eqphi}
\partial_{t}\partial_{z}\Phi&=&iA_{t}\partial_{z}\Phi+\frac{1}{2}[i\partial_{z}A_{t}\Phi+f\partial^{2}_{z}\Phi+f'\partial_{z}\Phi\nonumber\\
&\,&+(\partial-iA)^{2}\Phi-z\Phi],
\end{eqnarray}
\begin{equation}\label{constraint}
\partial_{z}(\partial_{z}A_{t}-\partial\cdot\boldsymbol{A})=i(\overline{\Phi}\partial_{z}\Phi-\Phi\partial_{z}\overline{\Phi}),
\end{equation}
\begin{eqnarray}\label{eqa}
\partial_{t}\partial_{z}\boldsymbol{A}&=&\frac{1}{2}[\partial_{z}(\boldsymbol{\partial} A_{t}+f\partial_{z}\boldsymbol{A})+(\partial^{2}\boldsymbol{A}-\partial\boldsymbol{\partial}\cdot\boldsymbol{A})\nonumber\\
&\,&-i(\overline{\Phi}\partial\Phi-\Phi\partial\overline{\Phi})]-\boldsymbol{A}\overline{\Phi}\Phi,
\end{eqnarray}
\begin{eqnarray}\label{eqat}
\partial_{t}\partial_{z}A_{t}&=&\partial^{2}A_{t}+f\partial_{z}\boldsymbol{\partial}\cdot\boldsymbol{A}-\partial_{t}\boldsymbol{\partial}\cdot\boldsymbol{A}-2A_{t}\overline{\Phi}\Phi\nonumber\\
&\,&+if(\overline{\Phi}\partial_{z}\Phi-\Phi\partial_{z}\overline{\Phi})-i(\overline{\Phi}\partial_{t}\Phi-\Phi\partial_{t}\overline{\Phi}).
\end{eqnarray}
The related numerical methods used in this paper include pseudo-spectral method and Runge-Kutta method. The pseudo-spectral method is used to represent the above functions with 25 Chebyshev modes in the $z$ direction and 241 Fourier modes in the $x,y$ direction. The Runge-Kutta method is used to evolve the equations in time direction with the time step $\delta t=0.05$. For a detail review of these methods, one can refer to Ref.\cite{chesler2013holographic,du2014holographic,2016jheplan,2016cqgguo}. While the initial bulk configurations for $\Phi, \boldsymbol{A}, A_{t}$ can be found in appendix A.

\section{numerical results}
\label{sec3}

\begin{figure}
\begin{center}
\includegraphics[scale=0.6]{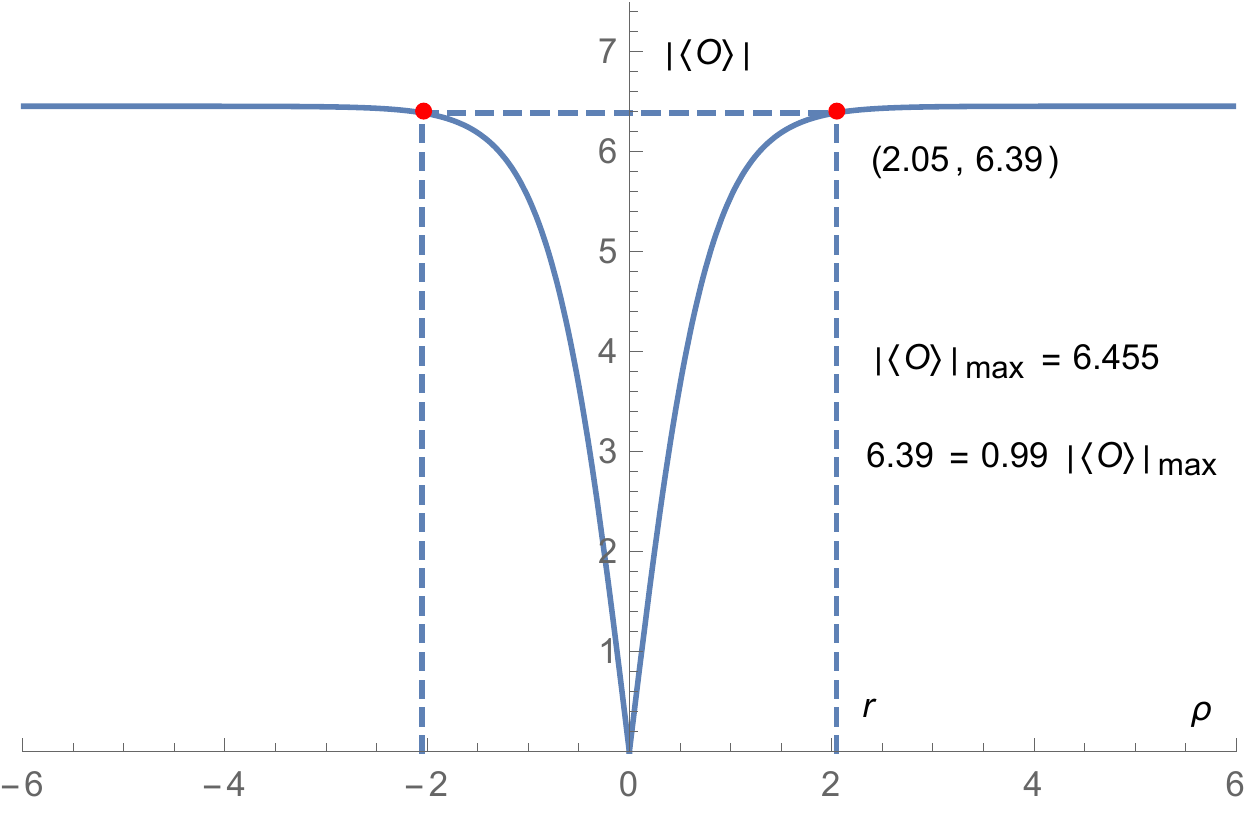}
\end{center}
\caption{The configuration of vortex for chemical potential $\mu=6$. We define the vortex radius as $r$, such that $|\langle O (r)\rangle|=0.99|\langle O \rangle|_{max}$. Here $r=2.05$.  }\label{vradius}
\end{figure}

Vortex and anti-vortex have the same configuration. The only difference is that one has clockwise rotation while the other has counterclockwise rotation. The static vortex solution can be obtained from the above equations of motion in the cylindrical coordinate\cite{2010prdkeranen2}. When the chemical potential of the superfluid is chosen as $\mu=6>\mu_{c}$, the vortex configuration is shown in Fig.\ref{vradius}. We define the vortex radius as $r$, such that $|\langle O (r)\rangle|=0.99|\langle O \rangle|_{max}$, where  $|\langle O \rangle|_{max}$ denotes the condensation of homogeneous superfluid solution. The vortex radius $r$ is found to be $2.05$.

\subsection{separation distance between vortex and anti-vortex}

\begin{figure}
\begin{center}
\includegraphics[scale=0.41]{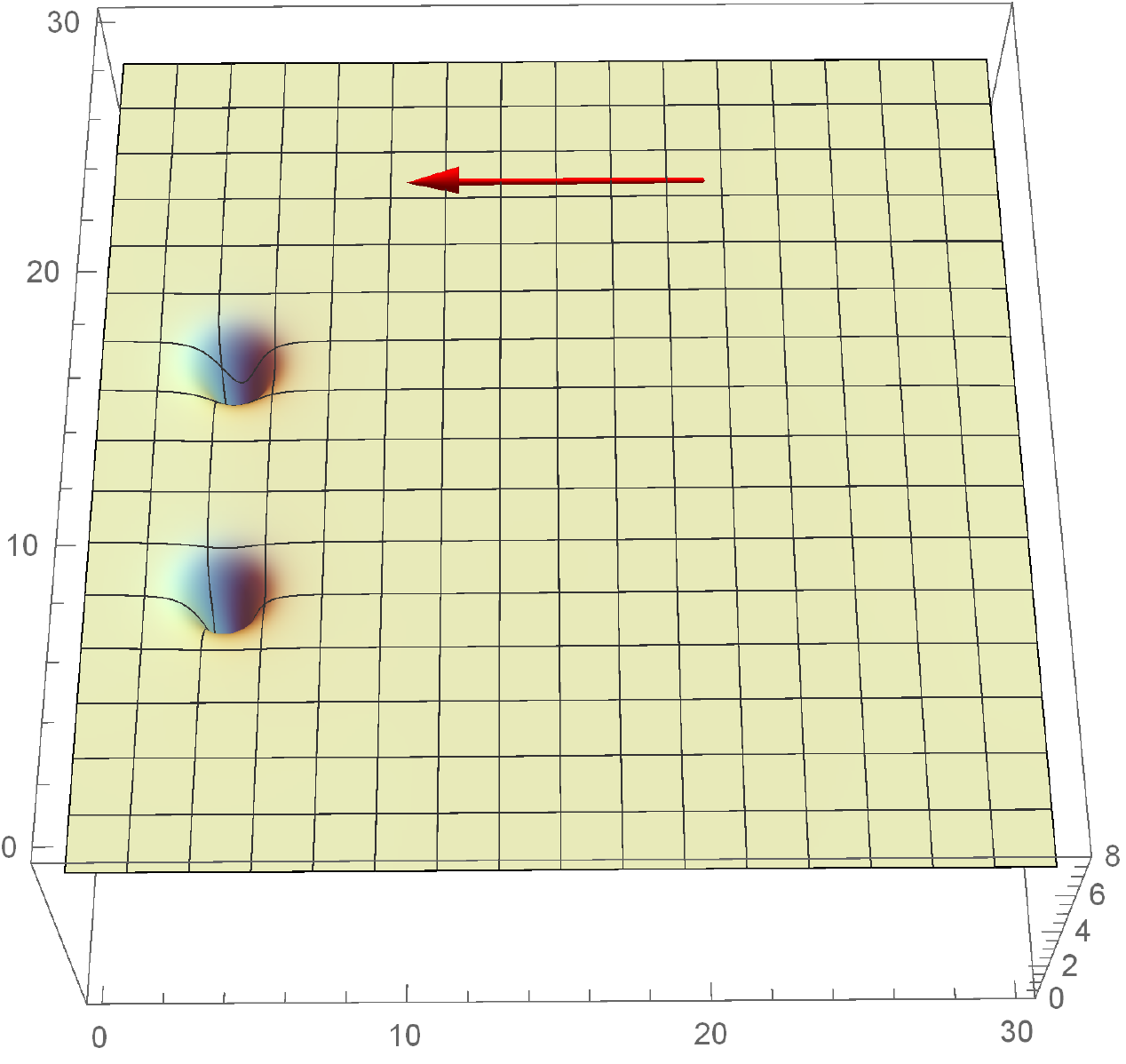}
\includegraphics[scale=0.47]{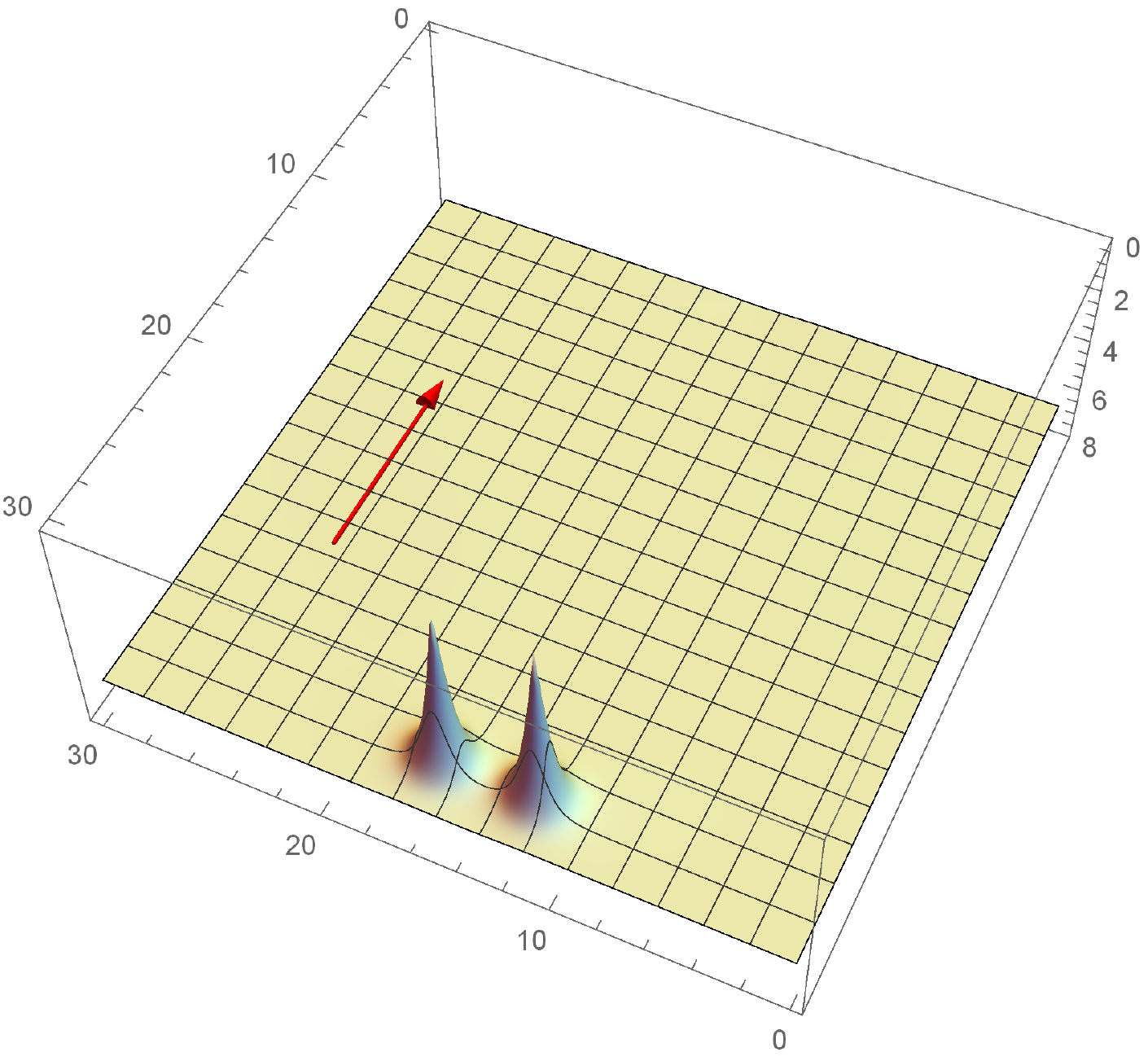}
\includegraphics[scale=0.44]{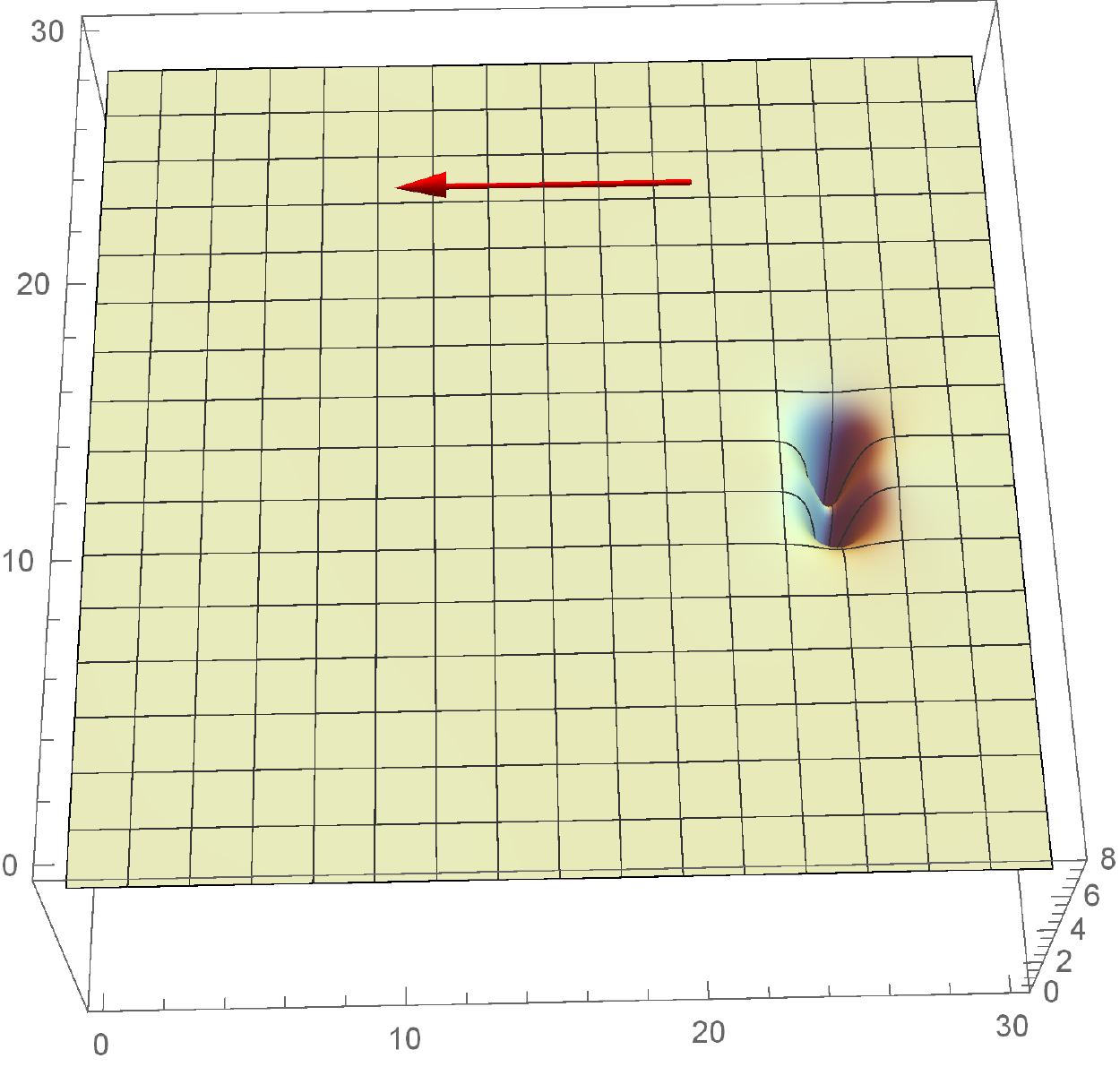}
\includegraphics[scale=0.43]{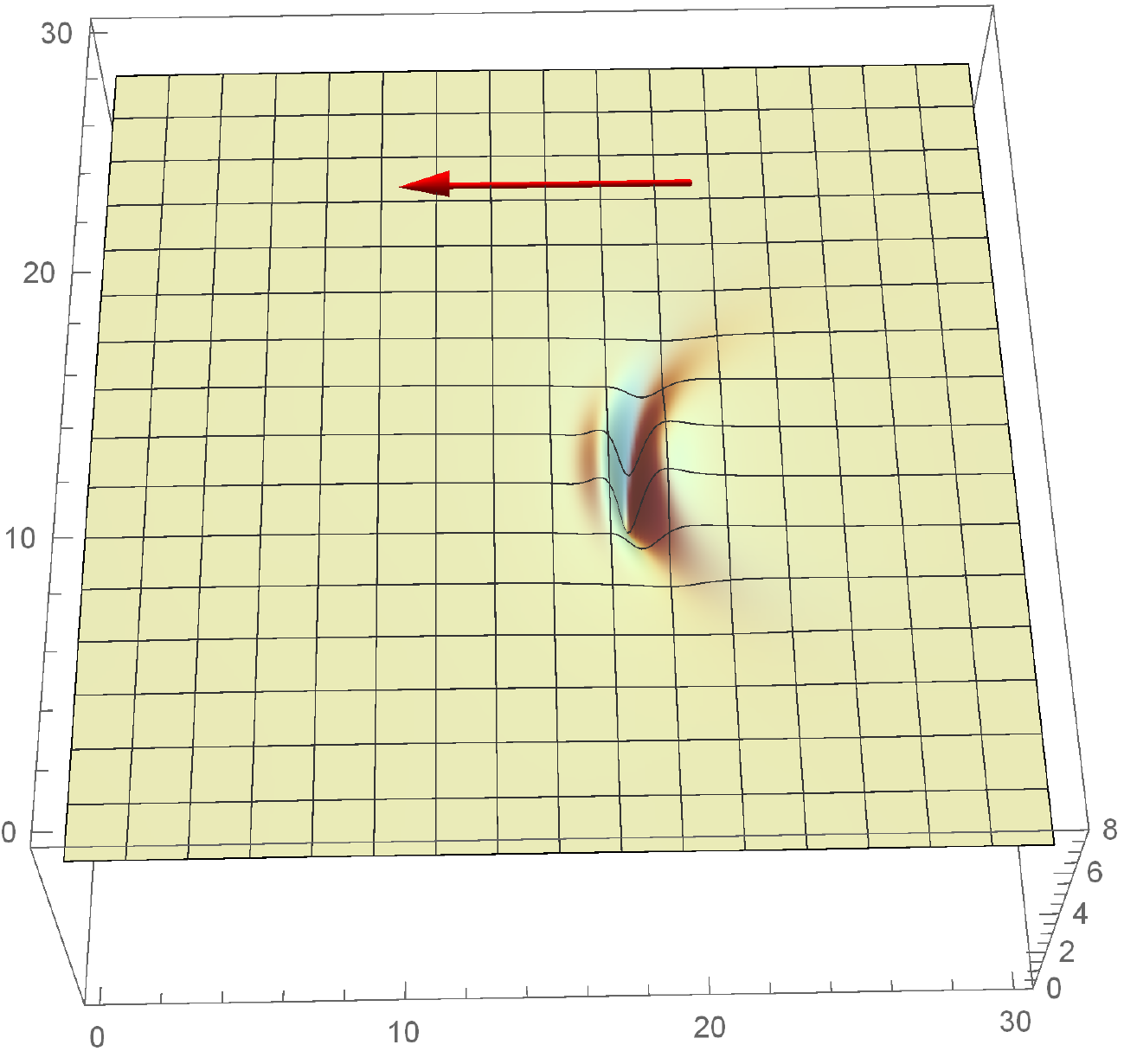}
\end{center}
\caption{The superfluid configurations for different time $t=400,592,640,660$. Red arrow indicates the moving direction. The separation distance $\delta(t)$ decreases with time until the vortices are annihilated. The graph is especially shown at time $t=592$, when the vortex and anti-vortex just touched, that is $\delta(t=592)=2r=4.1$.    }\label{vvst}
\end{figure}

We achieve long-time simulations of the vortex annihilation process and obtain 10 sets of data.  As examples, the superfluid configurations for different time $t=400,592,640,660$ are shown in Fig.\ref{vvst}. The graph is especially shown at time $t=592$, when the vortex and anti-vortex just touched, with $\delta(t=592)=2r=4.1$. The initial relative motion speed is small. The separation distance $\delta (t)$  decreases with time until the vortices are annihilated.

\begin{figure}
\begin{center}

\includegraphics[scale=0.5]{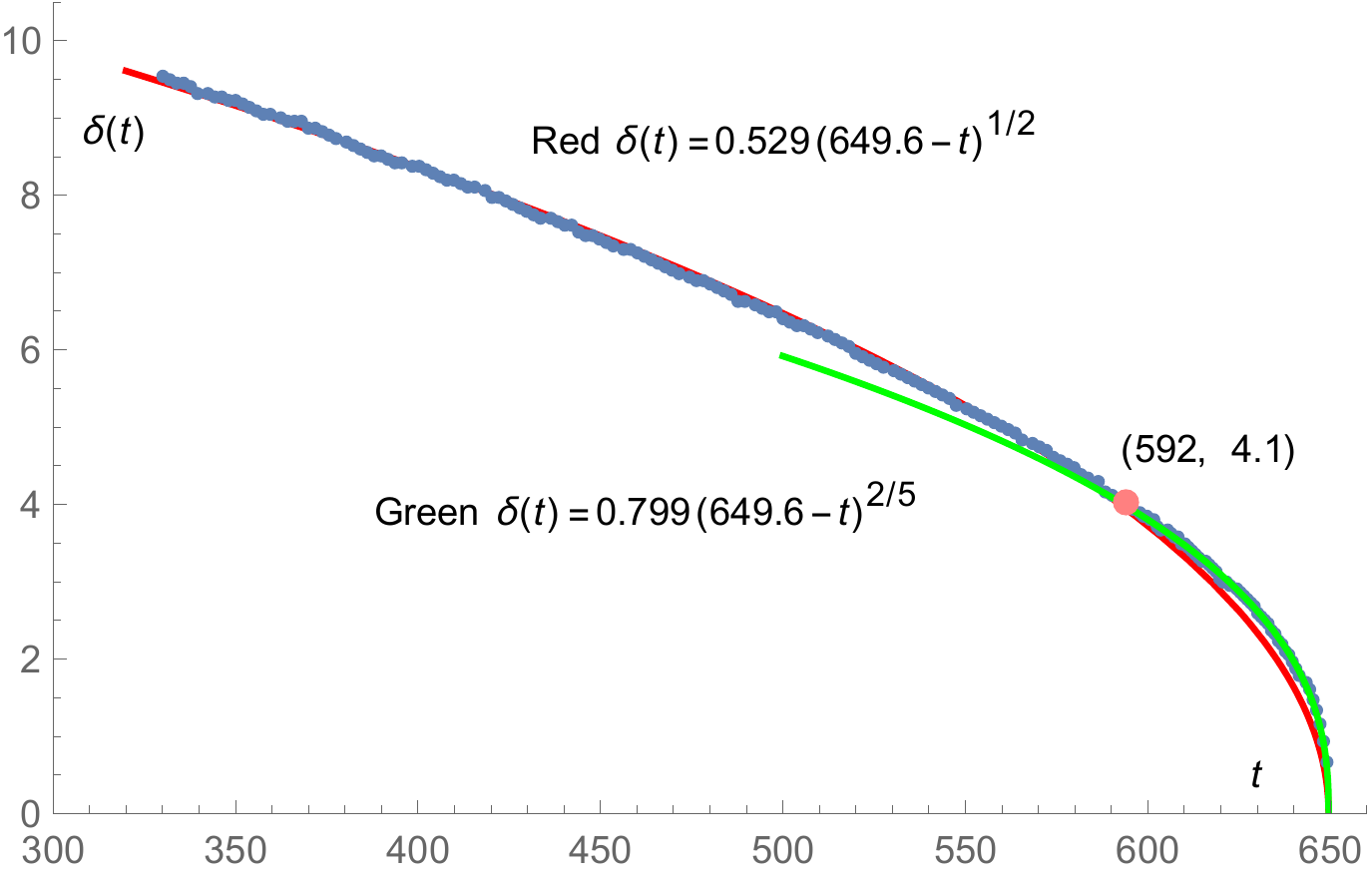}
\includegraphics[scale=0.5]{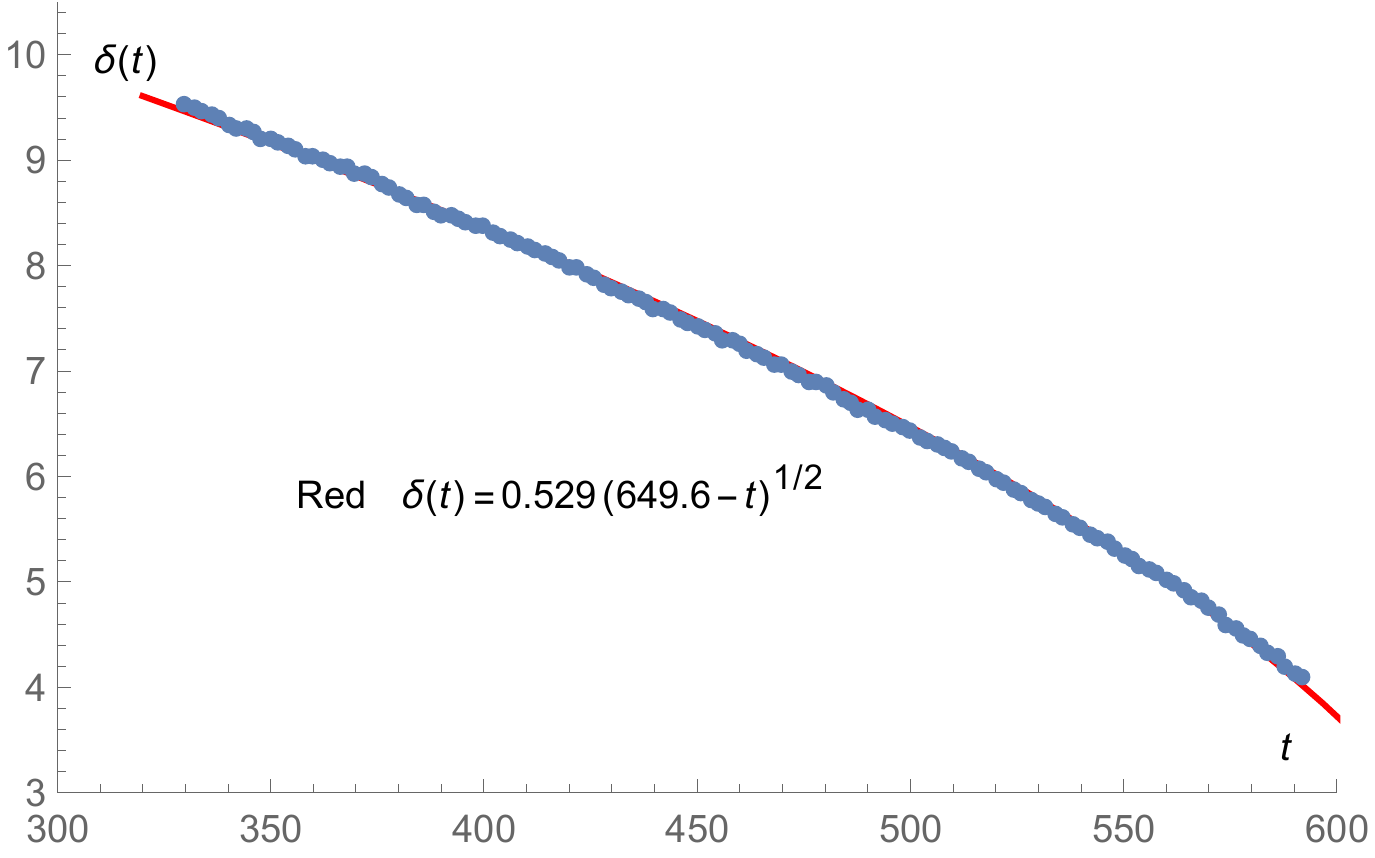}

\includegraphics[scale=0.5]{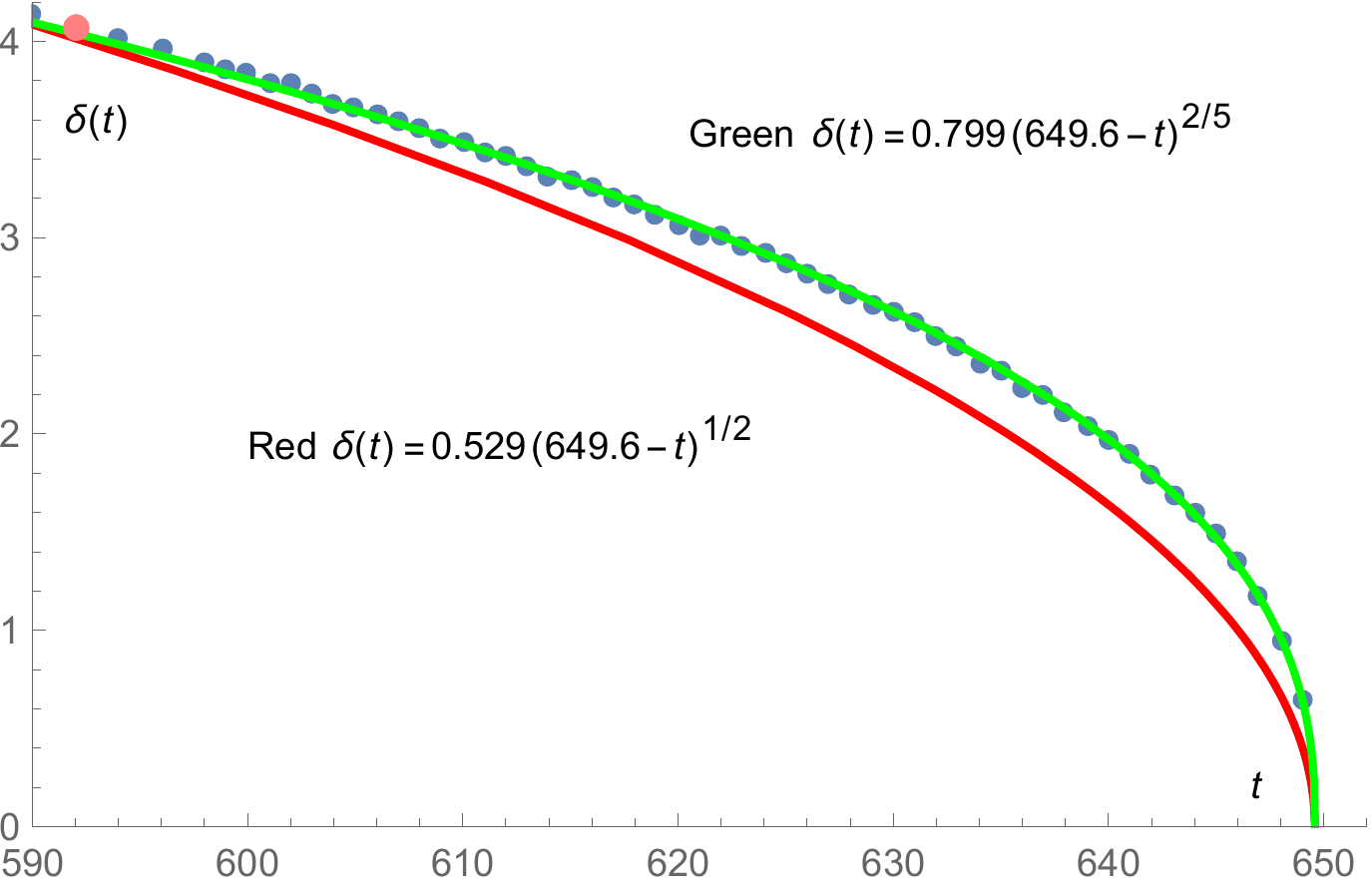}
\includegraphics[scale=0.5]{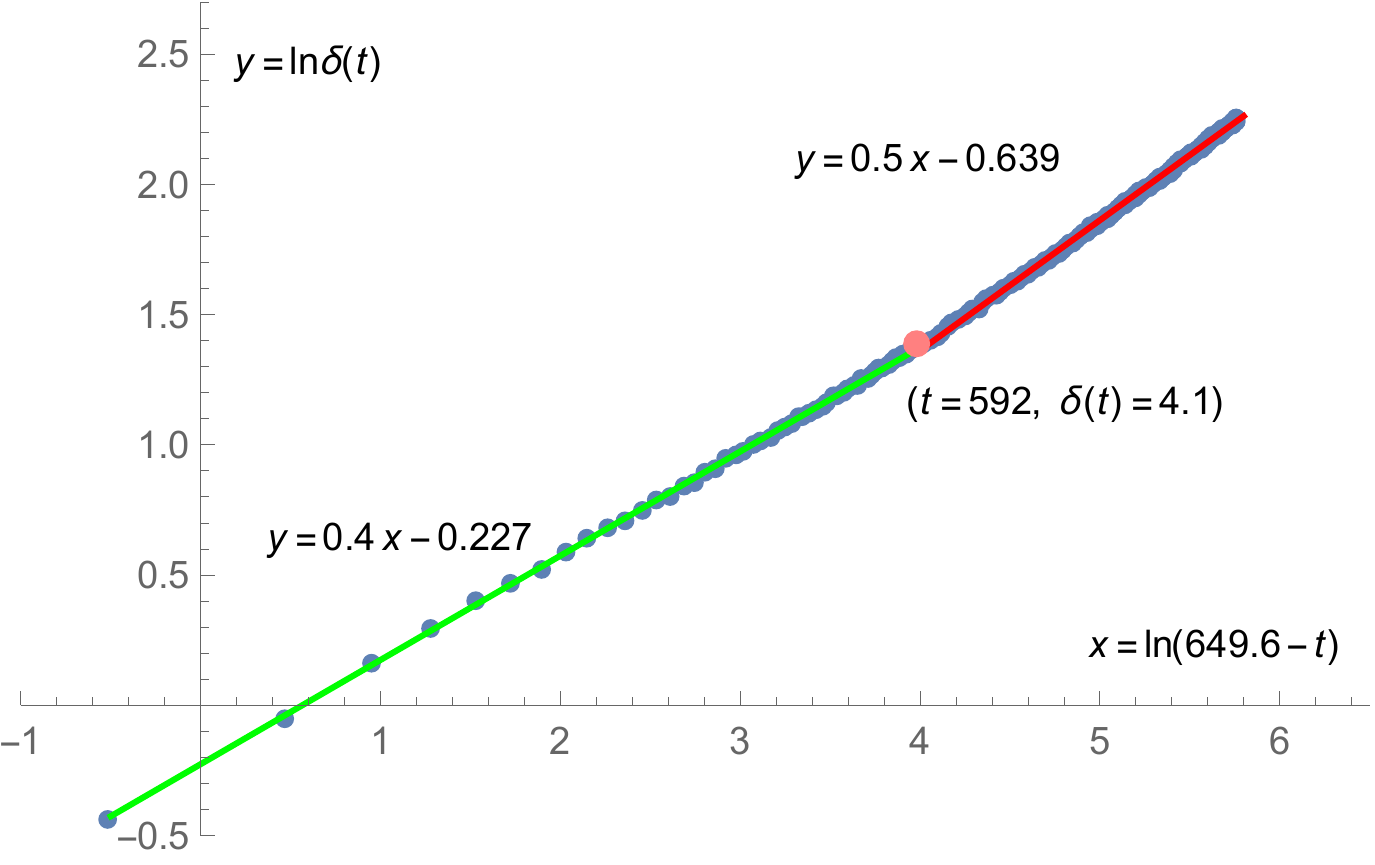}
\end{center}
\caption{The blue dots represent the real-time separation distance between the vortices. The solid red lines are the fitting curves ($\delta(t)=0.529(649.6 - t)^{1/2}$ and $y=0.5x-0.639$) for the first stage ($\delta(t)>4.1$), and the solid green lines are the fitting curves ($\delta(t)=0.799(649.6 - t)^{2/5}$ and $y=0.4x-0.227$) for the second stage($\delta(t)<4.1$). It clearly shows that the process can be divided into two stages, the first stage has a scaling exponent $1/2$ and the second stage has a scaling exponent $2/5$.  }\label{dist123}
\end{figure}

$\delta(t)$ is shown in Fig.\ref{dist123}. The blue dots represent the separation distance between the vortices at different times. We want to study the interaction between vortex and anti-vortex without being affected by other effects. So the separation distance is recorded every $\Delta t$ from time $t=330$, when the initially added perturbation modes are basically dissipated. From $t=330$ to $t=600$,  $\Delta t=2$. While for the rest of time, $\Delta t=1$. It can be seen from the figure that the separation distance of the vortices becomes smaller with time, and the relative motion between them becomes faster and faster, as if there is a mutually attractive force. In particular, from their contact $\delta(t)<2r=4.1$, the acceleration becomes larger, that is, the attractive force becomes larger. As a result, this annihilation process can be divided into two stages. The first part is the stage before the vortex pair contacts each other, while the second part is the stage after the contact. In Fig.\ref{dist123}, the solid red line is the fitting curve for the first stage ($\delta(t)>4.1$), and the solid green line is the fitting for the second stage($\delta(t)<4.1$).

As shown in the first three graphes, the first stage is well fitted by function
\begin{equation}\label{disat}
 \delta (t)=A (t_{0} - t)^{1/2},
\end{equation}
where $A=0.529$ is the fitting constant, $t_{0}=649.6$ is the time when the vortex pair annihilates.
The second stage is well fitted by function
\begin{equation}\label{disbt}
 \delta (t)=B (t_{0} - t)^{2/5},
\end{equation}
where $B=0.799$ is the fitting constant. The fourth graph is a log-log plot. The fitting function is
 \begin{equation}\label{fitfun2}
 ln(\delta(t))=n ln(t_{0}-t)+\varpi.
  \end{equation}
It clearly shows that the process can be divided into two stages, the first stage has a scaling exponent $n=0.5$ and the second stage has a scaling exponent $n=0.4$. The results of the fourth graph are consistent with those of the first three graphes. In the 10 sets of simulation experiments, the average value of $A$ is 0.529, and $B$ is 0.798. More detailed data analysis can be found in appendix B.

A comparison of our results with the results of previous work is interesting. First of all, the scaling law $1/2$ of the separation distance between vortices for $\delta(t)>2r$ seems to be universal for two dimensional and three dimensional cases. It is predicted by simple dimensional analysis\cite{1994prlwaele,2017prfalberto}. It is also confirmed by experiments\cite{2008PNASbewley,2008pdnppaoletti},and by an approximate analytic solution \cite{2003jltsergey}. Although some numerical studies report small modified scaling exponents\cite{2011jltpbarenhi,2012prbbaggaley, 2012pfzuccher, 2014praallen,2017prfalberto}.  Secondly, when $\delta(t)<2r$, a novel scaling exponent $2/5$ is obtained in this paper. Due to the small vortex radius, minor differences between the two scaling exponents $1/2$ and $2/5$ may have not yet been observed in other works.

\subsection{attractive interaction}

\begin{figure}
\begin{center}
\includegraphics[scale=0.4]{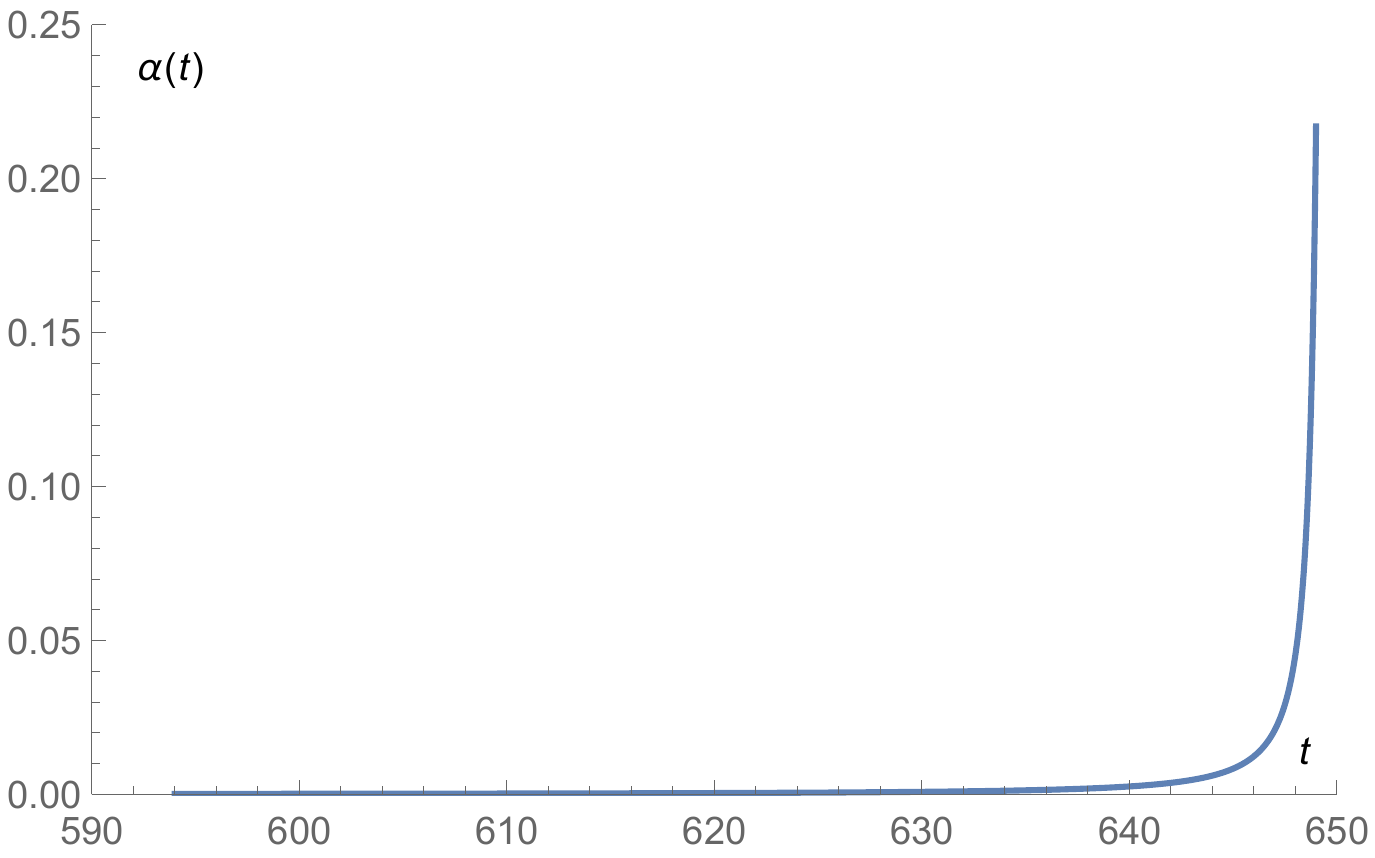}
\includegraphics[scale=0.4]{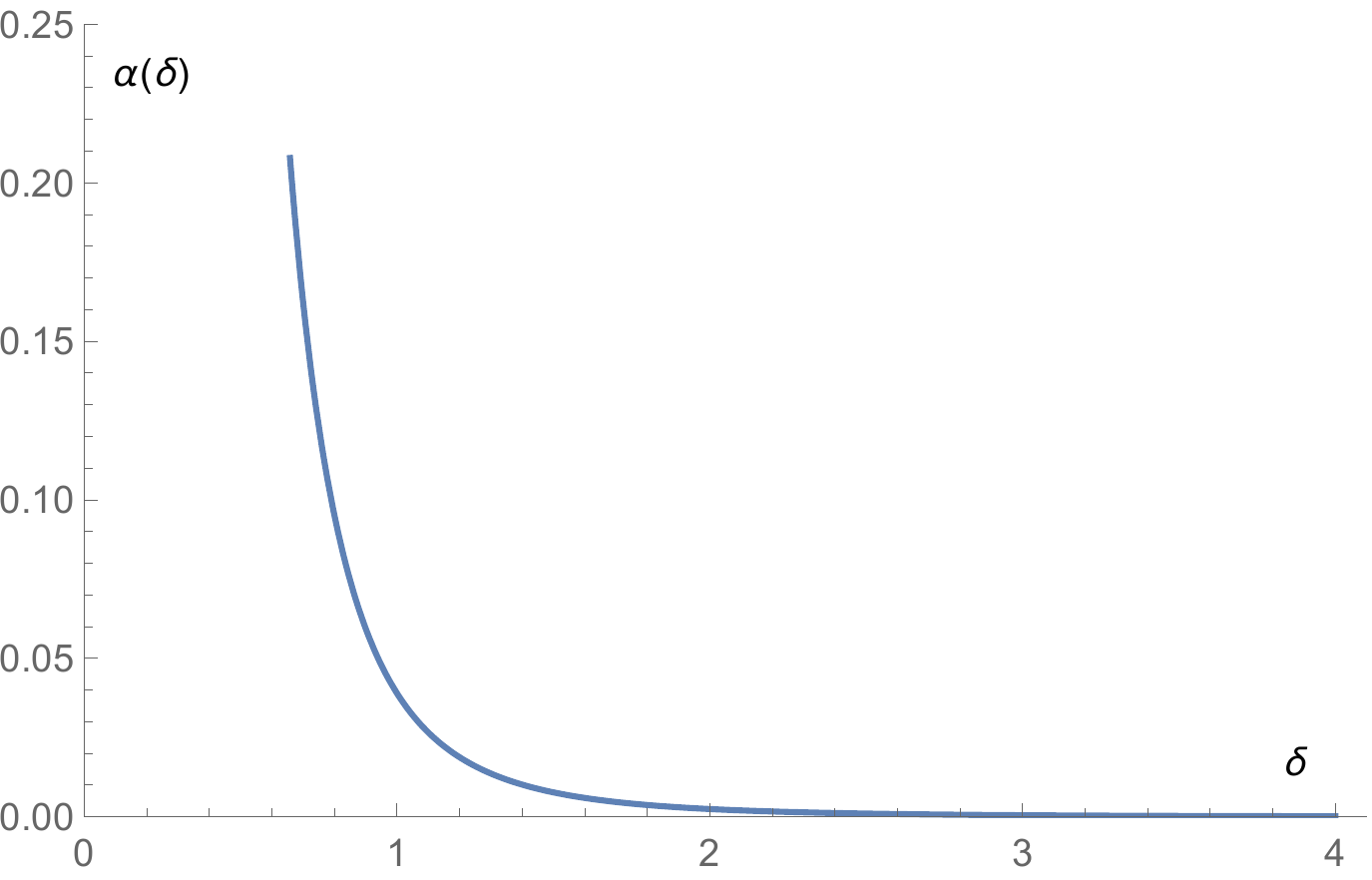}
\end{center}
\caption{For the second stage, left graph is $a(t)$, right graph is $a(\delta)$. When the separation distance is far away, the acceleration is extremely small.   }\label{atr12}
\end{figure}

Eq.(\ref{disat}) and Eq.(\ref{disbt}) can be rewritten into the following form,
\begin{equation}\label{fitfun}
  \delta(t)=\alpha (t_{0}-t)^{n}.
\end{equation}
Then the velocity that the vortices are approaching to each other is derived as
\begin{eqnarray}
  v(t)=\frac{d\delta(t)}{dt}= -n\alpha(t_{0}-t)^{n-1}.
\end{eqnarray}
The expression of velocity about distance can be obtained as
\begin{eqnarray}
  v(\delta)= -n\alpha^{\frac{1}{n}}\delta^{1-\frac{1}{n}}.
\end{eqnarray}
The expression of acceleration about time is derived as
\begin{eqnarray}
  a(t)=\frac{d^{2}\delta(t)}{dt^{2}}= n(n-1)\alpha(t_{0}-t)^{n-2}.
\end{eqnarray}
The expression of acceleration about distance can be obtained as
\begin{eqnarray}
  a(\delta)= n(n-1)\alpha^{\frac{2}{n}}\delta^{1-\frac{2}{n}}.
\end{eqnarray}
If the vortices are loosely treated as  particles, then according to the expression of the acceleration with respect to the separation distance, we can derive the attractive force between the vortex and anti-vortex as a function of separation distance,
\begin{eqnarray}
  f(\delta)\propto |a(\delta)|= |n(n-1)\alpha^{\frac{2}{n}}\delta^{1-\frac{2}{n}}|.
\end{eqnarray}
For the first stage, $n = 1/2$, the attractive force is
\begin{eqnarray}
  f(\delta>2r)\propto \frac{1}{\delta^{3}}.
\end{eqnarray}
For the second stage, $n = 2/5$, the attractive force is
\begin{eqnarray}
  f(\delta<2r)\propto  \frac{1}{\delta^{4}}.
\end{eqnarray}
For the second stage, the  acceleration as a function of time and as a function of separation distance is shown in Fig.\ref{atr12}. When the separation distance is far away,  the acceleration is extremely small, and the velocity increases very slowly.

\subsection{ vortex annihilation rate in superfluid turbulence with low vortex density}

It can be seen that only when the separation distance is very small, even less than the vortex size $ 2r = 4.1 $, the acceleration is larger. Here we present a theoretical interpretation  why the annihilation process of vortices in superfluid turbulence obeys the two-body decay law when the vortex density is low\cite{du2014holographic,2016jheplan,2018prabaggaley}. As the separation distance of vortices are large, the attractive force between them is small and the acceleration is small. Also because of the chaotic motion of superfluid turbulence, the effect of the long distance attractive force between vortex and anti-vortex is weakened. Therefore the average velocity of the vortex can be regarded as a constant $v_{0}$ for $\delta>2r$. So the annihilation probability for a pair of vortices should be proportional to the area ($v_{0}2r$) swept by one vortex per unit time, and inversely proportional to the area ($l^{2}/N(t)$) occupied by a single vortex. That is
\begin{equation}
  P_{1}\propto \frac{v_{0}2r}{l^{2}/N(t)}.
\end{equation}
Then the total annihilation rate should be
\begin{eqnarray}
  \frac{dN(t)}{dt}= -N(t)P_{1}\propto -N(t)\frac{v_{0}2r}{l^{2}/N(t)}=-\frac{2r v_{0}}{l^{2}}N(t)^{2},
\end{eqnarray}
or rewritten in terms of the vortex number density
\begin{eqnarray}
  \frac{dn(t)}{dt}\propto -2r v_{0} n(t)^{2}=-C n(t)^{2},
\end{eqnarray}
where $C$ is a constant. The above equation denotes the two-body decay law. $n(t)$ is obtained as
\begin{eqnarray}
  n(t)=\frac{1}{C t+1/n_{0}}\propto (t+\alpha)^{\beta},
\end{eqnarray}
where $\beta=-1$ which is successfully interpreted the results in Ref.\cite{du2014holographic,2016jheplan,2018prabaggaley}.

\section{conclusion and discussion}
\label{sec4}

In this paper, annihilation process of vortices in two-dimensional superfluid is numerically simulated based on holographic duality.  Firstly, the separation distance between vortex and anti-vortex as a function of time is recorded. The function is well fitted by $\alpha (t_{0}-t)^{n}$, where the scaling exponent $n=1/2$ for $\delta (t)>2r$, and $n=2/5$ for $\delta(t)<2r$. Thus the annihilation process can be divided into two stages which are separated by the vortex size $2r$.

Secondly, the approaching velocity as a function of time and as a function of separation distance are probed. When the separation distance is far away, the velocity increases very slowly and the acceleration is extremely small. If the vortices are loosely treated as  particles, we obtain the attractive force $f(\delta)\propto 1/\delta^{3}$ for the first stage, and $f(\delta)\propto 1/\delta^{4}$ for the second stage.

Thirdly, according to the characteristics of acceleration, we can reasonably assume that the average velocity of vortex in turbulent state is a constant when $\delta>2r$. Then the annihilation rate is derived as $ \frac{dn(t)}{dt}\propto -C n(t)^{2}$ which is called the two-body decay law. Thus we successfully explained why the annihilation process of vortices in superfluid turbulence obeys the two-body decay law when the vortex density is low\cite{du2014holographic,2016jheplan,2018prabaggaley}.

In the end, we would like to emphasize that the scaling exponent $1/2$ for $\delta(t)>2r$ in two-dimensional superfluid is same with many results\cite{1994prlwaele,2008PNASbewley,2008pdnppaoletti,2003jltsergey,2017prfalberto} obtained in three-dimensional cases. This means that the dimensional analysis argument of Ref.\cite{1994prlwaele,2017prfalberto} works also in this two-dimensional holographic superfluid case. The only relevant dimensional quantity is the circulation quantum. But their dimensionless parameters are different. What's more, another scaling exponent $2/5$ is obtained for $\delta(t)<2r$ in this paper. This new discovery may be examined by experiment with higher resolution. In this paper, we only consider the chemical potential $\mu=6$ case. The dependence of the power law on chemical potential will be revealed in our next paper.  All of the above results are based on holographic duality.  The two-dimensional cases in other models deserve to be disclosed in the future research. We will give the two-dimensional result within the GPE model in the near future. Investigation of the interaction between two same kind of vortices is also interesting. According to the symmetry, the interaction should be repulsive. When the separation distance is large, the repulsive behavior should be the same as the attractive force. When the separation distance is small, these two forces should behave differently. Anyway, the final result deserves to be revealed.

\acknowledgments
This research is supported by National Natural Science Foundation of China (Grant Nos.11847001,
11605082,11747017), Natural Science Foundation of Guangdong Province, China (Grant Nos.2016A030310363, 2016A030307051, 2015A030313789), Department of Education of Guangdong Province, China (Grant Nos.2017KQNCX124, 2017KZDXM056) and the Lingnan Normal University Project ZL1931.

\section*{ Appendix A: initial bulk configurations for $\Phi, \boldsymbol{A}, A_{t}$}
At chemical potential $\mu=6$, the static superfluid are randomly located a pair of vortex and anti-vortex and are added a random perturbation velocity field. To achieve the initial configuration, firstly the static superfluid solution\cite{20083h1} is multiplied with normalized vortex and anti-vortex solutions\cite{2010prdkeranen2}, e.g. $\Phi_{eq}(z)\frac{ \Phi_{vortice}(z,x-x_{1},y-y_{1})}{\Phi_{eq}(z)}\frac{ \Phi_{anti-vortice}(z,x-x_{2},y-y_{2})}{\Phi_{eq}(z)}$, where $(x_{1},y_{1})$ and $(x_{2},y_{2})$ are the coordinates of the vortex and anti-vortex. Secondly, based on Eq.(\ref{velocity}), the above obtained configuration $\Phi(z,x,y)$ can be multiplied with a random phase $e^{i \chi(x,y)}$ to add the perturbation velocity field. Here we take $\chi(x,y)= Re \gamma\Sigma_{k_{x}=-n_{x}\Lambda}^{n_{x}\Lambda}\Sigma_{k_{y}=-n_{y}\Lambda}^{n_{y}\Lambda}\xi(\boldsymbol{k})e^{i\boldsymbol{k}\cdot\boldsymbol{x}}$, where $\gamma$ is a small constant, $n_{x},n_{y}$ are small integers with $\Lambda=\frac{2\pi}{30}$, and $\xi(\boldsymbol{k})$ is a set of $\mathcal{O}(1)$ random complex coefficients\cite{2016jheplan}. In this paper, $\gamma$ is set to be $0.04$, $n_{x}$ and $n_{y}$ are taken as the same value with two cases as $4$ and $10$.

\section*{ Appendix B: data analysis of Fig.\ref{dist123}}

In Fig.\ref{dist123}, the blue dots $(t_{m},\delta _{m})$ represent the real-time separation distance  of the vortex pair. Data analysis start from $(t_{1}=330,\delta _{1}=9.535)$. When the separation distance is large, the attractive force is extremely small. Meanwhile, at early time, the initially added perturbation modes which are not basically dissipated have an effect on motion of vortices. As a result, behavior of $(t_{m},\delta _{m})$ before $t_{1}=330$ is not so regular and the data is omitted.  By applying the least square method, we obtain the red fitting curves for the first stage and the green fitting curves for the second stage.

For the first three graphes, the fitting function is Eq.(\ref{fitfun}), ie $ \delta(t)=\alpha (t_{0}-t)^{n}$ which has three parameters $\alpha, t_{0}, n$. In the enlarged view, it can be determined that the range of $t_{0}$ is $(649.4-649.8)$. Then we use the average relative deviation value to measure the quality of the fitting functions. The average relative deviation function is designed as
\begin{eqnarray}
  \varepsilon(\alpha,t_{0},n)=\frac{\sum_{m=i_{1}}^{m=i_{2}}(|\alpha (t_{0}-t_{m})^{n}-\delta_{m}|/\delta_{m})}{i_{2}-i_{1}+1}\times 100\%.
\end{eqnarray}
To determine $t_{0}$, $i_{1}$ is taken as 137 where $t_{137}=601$ and $i_{2}$ is taken as 185 where $t_{185}=649$.
\begin{table}
\begin{center}
\begin{tabular}{|c|c|c|c|c|c|}
 \hline
 $t_{0}$&649.4 & 649.5& 649.6& 649.7& 649.8  \\
\hline
 $\alpha$ & 0.804 & 0.793& 0.783& 0.772& 0.763  \\
\hline
 n & 0.399& 0.402& 0.406& 0.409 & 0.412   \\
 \hline
 $\varepsilon(t_{0})$ & 0.883\% & 0.675\%& 0.502\% & 0.517\% & 0.587\% \\
 \hline
\end{tabular}
\end{center}
\caption{For a giving specific $t_{0}$ and by applying the least square method, the fitting function $\delta(t)=\alpha (t_{0}-t)^{n}$ is obtained, ie the values of $\alpha,n$ are determined. Then the average relative deviation values $\varepsilon(t_{0})$ for different fitting functions are calculated.}\label{fitf1}
\end{table}
For a giving specific $t_{0}$ and by applying the least square method, the fitting function $\delta(t)=\alpha (t_{0}-t)^{n}$ is obtained, ie the parameters $\alpha,n$ are determined. Then the average relative deviation values $\varepsilon(t_{0})$ for different fitting functions can be calculated and are listed in Table.\ref{fitf1}. One can find that the $t_{0}=649.6$ case is the best as  the average relative deviation value $\varepsilon(t_{0}=649.6)=0.502\%$ is minimum. What's more, the approximation of the exponent $n\approx 0.4$ is very good. The error of the exponent for the second stage can be defined as
\begin{eqnarray}
  \eta_{12}=\frac{n(t_{0}=649.6)-0.4}{0.5-0.4}\times 100\%=\frac{0.406-0.4}{0.1}\times 100\%=6\%.
\end{eqnarray}
Thus we set $n=2/5$  and $t_{0}=649.6$. The corresponding fitting function is obtained as $\delta(t)=0.799 (649.6-t)^{2/5}$, which has an average relative deviation value of $\varepsilon=0.547\%$, which is very small. In this point of view, the data is well fitted by the function. For the first stage, $i_{1}$ is taken as 1 where $t_{1}=330$ and $i_{2}$ is taken as 132 where $t_{132}=592$. The average relative deviation values $\varepsilon(t_{0}=649.6,n)$ for different fitting functions $\delta(t)=\alpha (649.6-t)^{n}$ are listed in Table.\ref{fitf2}.
\begin{table}
\begin{center}
\begin{tabular}{|c|c|c|c|c|c|c|}
\hline
 n & 0.48& 0.49& 0.50& 0.505 & 0.51 & 0.52   \\
\hline
 $\alpha$ & 0.588 & 0.558& 0.529& 0.515 & 0.501& 0.475  \\
 \hline
 $\varepsilon(t_{0})$ & 1.110\% & 0.738\%& 0.565\% & 0.533\% &  0.539\% & 0.709\% \\
 \hline
\end{tabular}
\end{center}
\caption{In the first stage, for a giving $t_{0}=649.6$ and a specific $n$, by applying the least square method, the fitting function $\delta(t)=\alpha (649.6-t)^{n}$ is obtained, ie the parameter $\alpha$ is determined. Then the average relative deviation values $\varepsilon(t_{0}=649.6,n)$ for different fitting functions are calculated.}\label{fitf2}
\end{table}
One can find that the $n=0.505$ case is the best as  the average relative deviation value $\varepsilon(t_{0}=649.6,n=0.505)=0.533\%$ is minimum. While, we set $n=1/2$ for the first stage where the average relative deviation value $\varepsilon(t_{0}=649.6,n=0.5)=0.565\%$ which is also small. So the approximation of the exponent $n=1/2$ is very good. The error of the exponent for the first stage can be defined as
\begin{eqnarray}
  \eta_{11}=\frac{0.505-0.5}{0.5-0.4}\times 100\%=5\%.
\end{eqnarray}

The fourth graph is a log-log plot. The fitting function is Eq.(\ref{fitfun2}), ie $ ln(\delta(t))=n ln(649.6-t)+\varpi$ which has two parameters $n, \varpi$. By applying the least square method, the fitting functions are determined with $n=0.499, \varpi=-0.632$ for the first stage and with $n=0.402, \varpi=-0.232$ for the second stage. If we set the value of $n$ first, the fitting function will possess only one parameter $\varpi$. For the first stage, setting $n=1/2$, the fitting function is determined with $\varpi=-0.639$ and the corresponding average relative deviation value $\varepsilon(n=1/2,\varpi=-0.639)=0.281\%$ which is small. For the second stage, setting $n=2/5$, the fitting function is determined with $\varpi=-0.227$ and the corresponding average relative deviation value $\varepsilon(n=2/5,\varpi=-0.227)=0.587\%$ which is also small. So the functions are well fitted and the approximation of the exponents are very good. Then the error of the exponent for the first stage can be defined as
\begin{eqnarray}
  \eta_{21}=\frac{0.499-0.5}{0.5-0.4}\times 100\%=-1\%.
\end{eqnarray}
The error of the exponent for the second stage can be defined as
\begin{eqnarray}
  \eta_{22}=\frac{0.402-0.4}{0.5-0.4}\times 100\%=2\%.
\end{eqnarray}

\section*{ Appendix C: 10 sets of simulation experiments}

As discussed in Appendix A, the vortices are randomly located on superfluid where a random perturbation velocity field are added. Therefore, the 10 sets of simulations have different initial conditions and the results are shown in Table.\ref{tenruns} with vortices annihilation time $t_{0}$, parameter $A$ in fitting function $\delta(t)=A(t_{0}-t)^{1/2}$, scaling exponent error $\eta_{11}$, parameter $B$ in fitting function $\delta(t)=B(t_{0}-t)^{2/5}$ and scaling exponent error $\eta_{12}$.
\begin{table}
\begin{center}
\begin{tabular}{|c|c|c|c|c|c|c|c|c|c|c|}
\hline
  \,& 1& 2& 3 & 4 & 5 & 6 & 7 & 8 & 9 & 10   \\
\hline
 $t_{0}$ & 649.6& 506.2& 482.2 & 670.5 & 489.4 & 1240.4 & 926.5 & 518.8 & 468.2 & 625.2  \\
 \hline
A & 0.529& 0.528& 0.528 & 0.529 & 0.530 & 0.531 & 0.527 & 0.528 & 0.528 & 0.529 \\
 \hline
$\eta_{11}$ & 5\%& 2\%& 9\% & 6\% & 8\% & 3\% & 6\% & 3\% & 5\% & 4\% \\
 \hline
B & 0.799&  0.794& 0.802 & 0.796 & 0.795 & 0.797 & 0.797 & 0.801 & 0.798 & 0.797 \\
 \hline
$\eta_{12}$ & 6\%& 9\%& 9\% & 7\% & 8\% & 4\% & 1\% & 6\% & 8\% & 4\% \\
 \hline
\end{tabular}
\end{center}
\caption{Results of 10 sets of simulation experiments. $t_{0}$ is the vortices annihilation time, $A$ is the parameter  in fitting function $\delta(t)=A(t_{0}-t)^{1/2}$, $\eta_{11}$ is the  error of scaling exponent $n=1/2$, $B$ is the parameter  in fitting function $\delta(t)=B(t_{0}-t)^{2/5}$ and $\eta_{12}$ is the error of scaling exponent $n=2/5$.}\label{tenruns}
\end{table}


\end{document}